\documentclass[conference]{IEEEtran}
\IEEEoverridecommandlockouts
\usepackage{cite}
\usepackage{amsmath,amssymb,amsfonts,amsthm}
\usepackage{algorithmic}
\usepackage{graphicx}
\usepackage{textcomp}
\usepackage{xcolor}
\PassOptionsToPackage{hyphens}{url}
\usepackage{hyperref}
\usepackage{url}
\usepackage{enumitem}
\usepackage[ruled,vlined]{algorithm2e}
\def\BibTeX{{\rm B\kern-.05em{\sc i\kern-.025em b}\kern-.08em
    T\kern-.1667em\lower.7ex\hbox{E}\kern-.125emX}}

\newtheorem{definition}{Definition}
\newtheorem{remark}{Remark}
\newtheorem{theorem}{Theorem}
\newtheorem{lemma}{Lemma}

\SetKwInOut{Input}{Input}
\SetKwInOut{Output}{Output}

\begin{document}

\title{Privacy-Preserving Biometric Matching Using Homomorphic Encryption*\\
{\footnotesize \textsuperscript{*}This paper is a corrected and extended version of \cite{GP_CM_TrustCom}.}
\thanks{Supported by the Luxembourg National Research Fund (FNR) (12602667).}
}

\author{\IEEEauthorblockN{Gaëtan Pradel}
\IEEEauthorblockA{\textit{INCERT}, Luxembourg \\
\textit{Royal Holloway, University of London}\\
Egham, United Kingdom \\
gpradel@incert.lu}
\and
\IEEEauthorblockN{Chris Mitchell}
\IEEEauthorblockA{\textit{Royal Holloway, University of London}\\
Egham, United Kingdom \\
me@chrismitchell.net}
}

\maketitle

\begin{abstract}
Biometric matching involves storing and processing sensitive user information.
Maintaining the privacy of this data is thus a major challenge, and homomorphic encryption offers a possible solution.
We propose a privacy-preserving biometrics-based authentication protocol based on fully homomorphic encryption, where the biometric sample for a user is gathered by a local device but matched against a biometric template by a remote server operating solely on encrypted data.
The design ensures that 1) the user's sensitive biometric data remains private, and 2) the user and client device are securely authenticated to the server.
A proof-of-concept implementation building on the TFHE library is also presented, which includes the underlying basic operations needed to execute the biometric matching.
Performance results from the implementation show how complex it is to make FHE practical in this context, but it appears that, with implementation optimisations and improvements, the protocol could be used for real-world applications.
\end{abstract}


\section{Introduction}
This paper proposes a privacy-preserving biometric-based authentication protocol based on fully homomorphic encryption (FHE), designed for use in the case where the biometric sample for a user is gathered by a local device but matched against a biometric template by a remote server.
The goal is to enable this to occur without the remote server,modeled as a honest-but-curious adversary,gaining access to any of the sensitive biometric data.
The privacy-preserving and authentication properties of the protocol are formally established.
A proof-of-concept C/C++ implementation building on the TFHE library due to Chillotti et al.\ \cite{DBLP:journals/joc/ChillottiGGI20} has also been developed, in which face matching is used as the biometric.
Performance results from this implementation are presented.
The results of the implementation confirm the difficulty of making FHE practical in such a scenario, but we suspect that, with optimisations and improvements, the protocol could be used for real-world applications.

As part of the proof-of-concept, all the elementary operations necessary to execute the protocol using FHE were implemented.
Thus, as a side contribution, we have provided a set of elementary arithmetic routines in the ciphertext domain\footnote{The implementation is hosted here: \url{https://github.com/lab-incert/threats}.}, which could be useful for other prototype implementations.

\paragraph{Homomorphic encryption}
Homomorphic encryption allows one to perform computations on encrypted data, without ever
decrypting it. This enables users to perform operations in untrusted environments. The idea of
performing computations on encrypted data was introduced in $1978$ by Rivest, Shamir and Adleman
\cite{DBLP:journals/cacm/RivestSA78}. While many homomorphic schemes have been proposed
\cite{DBLP:conf/tcc/BonehGN05,DBLP:conf/ccs/NaccacheS98,DBLP:conf/eurocrypt/OkamotoU98,DBLP:conf/eurocrypt/Paillier99},
it wasn't until $2009$ that Gentry presented \cite{homenc} the first FHE scheme, based on ideal
lattices. Gentry's breakthrough rests on a technique called \emph{bootstrapping}. An FHE scheme
based on Gentry's blueprint enables an arbitrary number of additions and multiplications, i.e.\ any
function, to be computed on encrypted data. Since then, many other schemes have been proposed
\cite{DBLP:conf/focs/BrakerskiV11,DBLP:conf/innovations/BrakerskiV14,DBLP:conf/eurocrypt/DijkGHV10,DBLP:conf/crypto/GentryHS12,DBLP:conf/pkc/SmartV10,DBLP:conf/asiacrypt/StehleS10},
including schemes not using the bootstrapping technique. For example, in 2012, Brakerski, Gentry
and Vaikuntanathan\cite{DBLP:conf/innovations/BrakerskiGV12} presented a scheme based on the
ring version of the Learning With Errors problem, introduced by Regev
\cite{DBLP:conf/stoc/Regev05}. A second type of FHE scheme was introduced by Gentry, Sahai and
Waters \cite{DBLP:conf/crypto/GentrySW13}. This scheme was further improved
\cite{DBLP:conf/crypto/Alperin-SheriffP14,DBLP:conf/eurocrypt/DucasM15}, and most recently by
Chillotti et al. \cite{DBLP:conf/asiacrypt/ChillottiGGI16,DBLP:journals/joc/ChillottiGGI20}.

\paragraph{Biometric authentication}
The use of biometrics for authentication has been discussed for several decades, and has seen growing use.
International organisations suggest passwordless\footnote{See for example the World Economic Forum: \url{https://www.weforum.org/agenda/2020/04/covid-19-is-a-reminder-that-its-time-to-get-rid-of-passwords/}.} systems for authentication, and biometrics can solve this issue.
Advances mean that in some circumstances biometric recognition algorithms perform better than humans, even for face recognition \cite{DBLP:conf/aaai/LuT15}.
Nonetheless, biometric authentication faces a range of challenges \cite{DBLP:journals/scn/PagninM17}, in particular regarding the protection of users' sensitive data.
Biometric data, such as a fingerprint, is fixed for a lifetime, meaning that its use gives rise to significant privacy concerns.
Ideally, biometric data should not be processed without protection or anonymisation.
Homomorphic encryption offers a possible solution to this problem \cite{DBLP:journals/scn/PagninM17}, as it allows the authentication provider to perform biometric matching on (encrypted) data, while protecting the privacy of sensitive biometric data.

\paragraph{Related work}
The use of homomorphic cryptography in the context of biometric matching is not new
\cite{DBLP:conf/sp/OsadchyPJM10,DBLP:conf/eurocrypt/SchoenmakersT06}. However, most previous work
uses partially homomorphic encryption and not FHE\@.
Some of this work has promising performance results, e.g.\ Blanton and Gasti
\cite{DBLP:conf/esorics/BlantonG11} who calculate the Hamming distance between two iris feature
vectors in only 150 ms. However, because of the additive-only (partially homomorphic)
characteristic of the encryption schemes they use, they are not able to evaluate a circuit much
more complex than for Hamming distance. Yasuda et al.\ \cite{DBLP:conf/IEEEares/YasudaSKYK13} used
a homomorphic scheme that also enables multiplications in the ciphertext domain, but still only
compute the Hamming distance between two biometric vectors; moreover, the approach is vulnerable
against malicious attackers \cite{AbidinPK14}. Back in 2008, Bringer and Chabanne
\cite{DBLP:conf/africacrypt/BringerC08} proposed an authentication protocol based on the
homomorphic properties of two partially homomorphic encryption schemes.

Biometric matching based on FHE has been previously proposed; perhaps the first example is the
private face verification system of Troncoso-Pastoriza et al.\ \cite{DBLP:journals/tifs/Troncoso-PastorizaGP13}.
Cheon et al.\ \cite{DBLP:journals/iacr/CheonCKL16} proposed \emph{Ghostshell}, a tool that works on iris templates, that is computationally costly.
More recently, Boddeti \cite{DBLP:conf/btas/Boddeti18} showed how to execute a secure face matching using the Fan-Vercauteren FHE scheme \cite{DBLP:journals/iacr/FanV12} and obtained practical results by packing the ciphertexts in a certain way.

\paragraph{Structure of the paper}
Section~\ref{sec:prelim} introduces the notions necessary for the rest of the paper.
Sections~\ref{sec:contrib} and~\ref{sec:security} are the core of the paper, presenting the design
and security properties of the protocol. Finally, Sections~\ref{sec:implementation}
and~\ref{sec:conclu} give results from the protocol implementation and conclude the paper.

\section{Preliminaries}
\label{sec:prelim}

$\mathbb{N}$, $\mathbb{Z}$, $\mathbb{R}$ and $\mathbb{B}$ represent the sets of natural numbers, integers, reals and bits, respectively.

\subsection{Security notions}
\label{subsec:sec}

We next introduce some formal security notions.
For more complete versions of Definitions~\ref{def:negligible}-\ref{def:statistical_indistinguishability}, see Goldreich~\cite{DBLP:books/cu/Goldreich2001}.

\begin{definition}[Negligible]
\label{def:negligible}
We say a function $f : \mathbb{N} \mapsto \mathbb{R}$ is \emph{negligible} if for every polynomial $p$ there exists an $N$ such that, for all $ n > N$:
\[ f(n) < \frac{1}{p(n)} .\]
\end{definition}

\begin{definition}[Probability ensemble]
\label{def:probability_ensemble}
Let $I$ be a countable index set.
A \emph{probability ensemble} indexed by I is a sequence of random variables indexed by $I$.
Namely, any $X = (X_i )_{i \in I},$ where each $X_i$ is a random variable, is an ensemble indexed by $I$.
\end{definition}


\begin{definition}[Polynomial-time indistinguishability]
\label{def:computionally_indistinguishability} Suppose $X = ( X_i )_{i \in \mathbb{N}}$ and $Y =
(Y_i )_{i \in \mathbb{N}}$ are ensembles with index set $\mathbb{N},$ where $X_i, Y_i \in
\mathbb{B}^n$ for all $i$. Then $X$ and $Y$ are said to be \emph{indistinguishable in
polynomial-time} if, for every probabilistic polynomial-time algorithm $D: \mathbb{B}^n \rightarrow
\mathbb{B}$, every polynomial $p$, and all sufficiently large $n$:
\[ |\text{Pr}[D(X_n) = 1 ] - \text{Pr} [D(Y_n) = 1] | < \frac{1}{p(n)}.\]

We write $X \overset{c}{\approx} Y.$
\end{definition}

\begin{remark}
\label{rmk:computationally_indistinguishability}

We follow common practice and refer to \emph{computational} \emph{indistinguishability} instead of indistinguishability in polynomial-time.
\end{remark}

\begin{definition}[Statistical distance]
\label{def:statistical_distance}
Suppose $X = ( X_i )_{i \in \mathbb{N}}$ and  $Y = ( Y_i )_{i \in \mathbb{N}}$ are ensembles with index set $\mathbb{N},$ where $X_i, Y_i \in \mathbb{B}^n$ for all $i$.
Then the \emph{statistical distance} function $\Delta \colon \mathbb{N} \rightarrow \mathbb{R}$  is defined as:
\[ \Delta(n) \overset{\text{def}}{=} \frac{1}{2} \sum \limits_{\alpha \in \mathbb{B}^n} |\text{Pr}[X_n = \alpha] - \text{Pr}[Y_n = \alpha]| .\]
\end{definition}

\begin{definition}[Statistical indistinguishability]
\label{def:statistical_indistinguishability}
Suppose $X = ( X_i )_{i \in \mathbb{N}}$ and  $Y = ( Y_i )_{i \in \mathbb{N}}$ are ensembles with index set $\mathbb{N},$ where $X_i, Y_i \in \mathbb{B}^n$ for all $i$.
Then $X$ and $Y$ are said to be \emph{statistically indistinguishable} if their statistical distance is negligible.

We write $X \overset{s}{\approx} Y$.
\end{definition}

\begin{remark}
\label{rmk:statistical_indistinguishability}
If the ensembles $X$ and $Y$ are statistically indistinguishable, then they are also computationally indistinguishable. The converse is not true.
\end{remark}

\begin{definition}[Adversary]
\label{def:adversary}
An \emph{adversary} $\mathcal{A}$ for a cryptographic scheme is a polynomial-time algorithm (or a set of polynomial-time algorithms) that models a real-world attacker.
It is equipped with defined computational resources and capabilities, and is designed to attack the security of the scheme typically as a participant in a security game.
\end{definition}

\begin{definition}[Challenger]
\label{def:challenger}
A \emph{challenger} for a cryptographic scheme is a poly\-nomial-time algorithm (or a set of polynomial-time algorithms) that models a real-world instance of the scheme.
It is usually assumed to possess unlimited computational resources and capabilities, and is viewed as a `black box' which responds to queries made by an adversary in a security game.
\end{definition}

\begin{definition}[Security game]
\label{def:security_game}
A \emph{security game} models an attack on a cryptographic scheme involving an adversary and a challenger.
\end{definition}

\begin{definition}[Advantage]
\label{def:advantage}
In the context of a cryptographic scheme and a security game for this scheme, the \emph{advantage} of an adversary is a function of the probability that the adversary wins the security game that measures the adversary's improvement over random choice.
\end{definition}

\subsection{Homomorphic Encryption}
\label{sec:he}

We next formally introduce homomorphic encryption and certain associated notions.
For more complete versions of these definitions, see Armknecht et al.\ \cite{cryptoeprint:2015:1192}.

\begin{definition}[Homomorphic Encryption scheme]
\label{def:homomorphic_encryption} A \emph{homomorphic encryption scheme} $\mathcal{E}$ for a
circuit family $\Pi$ consists of four PPT algorithms
$(\mathsf{KeyGen}$, $\mathsf{Enc}$, $\mathsf{Dec}$, $\mathsf{Eval})$ with the following properties.
\begin{itemize}
\item $(sk, pk, evk) \leftarrow \mathsf{KeyGen}(1^{\lambda})$.
Given the security parameter $\lambda \in \mathbb{N}$, $\mathsf{KeyGen}$ outputs a key triple made up of a secret key $sk$, a public key $pk$ and an evaluation key $evk$.
The plaintext space $\mathcal{M}$ and the ciphertext space $\overline{\mathcal{M}}$ are determined by $pk$.
\item $\overline{m} \leftarrow \mathsf{Enc}(pk, m).$
Given a public key $pk$ and a plaintext $m \in \mathcal{M}$, $\mathsf{Enc}$ outputs a ciphertext $\overline{m} \in \overline{\mathcal{M}}$.
\item $\begin{cases} m \\ \perp \end{cases} \leftarrow \mathsf{Dec}(sk, \overline{m}).$
Given a secret key $sk$ and a ciphertext $\overline{m}$, $\mathsf{Dec}$ outputs either the plaintext $m \in \mathcal{M}$ if $\overline{m} \leftarrow \mathsf{Enc}(pk, m)$ or $\perp$.
\item $ \overline{m}' \leftarrow \mathsf{Eval} (evk, \pi, \overline{m}).$ Given an evaluation key $evk$, a circuit $\pi \in \Pi$, where $\Pi$ is a circuit family (see Appendix~\ref{apd:circuits} for details)  and a ciphertext $\overline{m} \in \overline{\mathcal{M}}$, $\mathsf{Eval}$ outputs another ciphertext $\overline{m}' \in \overline{\mathcal{M}}$.
\end{itemize}
\end{definition}

\begin{remark}
\label{rmk:keys}
Depending on the scheme, the evaluation key $evk$ might be part of, or equal to, the public key $pk$.
\label{rmk:circuit}
For simplicity of presentation, here and throughout we assume that the circuit input to $\mathsf{Eval}$ has input size corresponding to the size of the input ciphertext(s).
\end{remark}

Definition~\ref{def:homomorphic_encryption}, and those below, holds for a range of types of plaintext, including both bit strings and vectors of plaintexts.
Some algorithms, such as $\mathsf{KeyGen}$, take as input a security parameter $\lambda$, which will be denoted as such throughout this paper unless stated otherwise.
This input is usually written in unary representation $1^{\lambda}$ because we want an algorithm that runs in time polynomial in the size of $\lambda$ to be considered as efficient.
We refer to the outputs of $\mathsf{Enc}$ as `fresh ciphertexts' and those of $\mathsf{Eval}$ as `evaluated ciphertexts'.

\begin{definition}[Correctness]
\label{def:correct} Suppose $\mathcal{E}$ $=$  $(\mathsf{KeyGen}$, $\mathsf{Enc}$, $\mathsf{Dec}$, $\mathsf{Eval})$ is a homomorphic encryption scheme with security parameter $\lambda$\@.
We say $\mathcal{E}$ is  \emph{correct} for a circuit family $\Pi$ if $\mathcal{E}$ correctly
decrypts both fresh and evaluated ciphertexts, namely, for all $\lambda \in \mathbb{N}$, the
following two conditions hold.
\begin{itemize}
\item Suppose $(sk, evk, pk) \leftarrow \mathsf{KeyGen}(1^{\lambda})$.
If $m \in \mathcal{M}$ and $\overline{m} \leftarrow \mathsf{Enc}(pk, m)$ then $m \leftarrow \mathsf{Dec}(sk, \overline{m}).$
Else $\perp \leftarrow \mathsf{Dec}(sk, \overline{m}).$
\item For any key triple $(sk, evk, pk) \leftarrow \mathsf{KeyGen}(1^{\lambda})$, any circuit $\pi \in \Pi$, any plaintext $m \in \mathcal{M}$ and any ciphertext $\overline{m} \in \overline{\mathcal{M}}$ with $\overline{m} \leftarrow \mathsf{Enc}(pk, m)$,  if $\overline{m}' \leftarrow \mathsf{Eval}(evk, \pi, \overline{m})$ then $\mathsf{Dec}(sk, \overline{m}') \rightarrow \pi(m)$.
\end{itemize}
\end{definition}

\begin{definition}[Indistinguishability under Chosen-Plaintext Attacks security game]
\label{def:cpagame} Suppose $\mathcal{E}=(\mathsf{KeyGen}, \mathsf{Enc}, \mathsf{Dec},
\mathsf{Eval})$ is a homomorphic encryption scheme with security parameter $\lambda$. Suppose also
that $\mathcal{A}$ is a PPT adversary. The \emph{indistinguishability under chosen-plaintext
attacks} (\emph{IND-CPA}) \emph{security game} is as follows.
\begin{enumerate}
\item A challenger runs $(sk, pk, evk)  \leftarrow  \mathsf{KeyGen}(1^{\lambda})$ and shares $pk$ with $\mathcal{A}$.
\item $\mathcal{A}$ generates two distinct plaintexts $\{m_0, m_1\}$ and submits a query to the challenger to request the encryption of one of them with $pk$.
\item The challenger chooses $i \in \mathbb{B}$ uniformly at random, computes $\overline{m} \leftarrow \mathsf{Enc}(m_i, pk)$ and sends $\overline{m}$ to $\mathcal{A}$.
\item $\mathcal{A}$ outputs a pair $(m_j, \overline{m})$, where $j \in \mathbb{B}$, and wins the game if $i = j$.
\end{enumerate}
We denote this security game by $\text{IND-CPA}^{\mathcal{A}}_{\mathcal{E}} (1^{\lambda})$ and a win in an instance of this security game by $\text{IND-CPA}^{\mathcal{A}}_{\mathcal{E}} (1^{\lambda}) = 1$.
\end{definition}

\begin{definition}[Advantage for the IND-CPA security game]
\label{def:adv_INDCPA} Suppose $\mathcal{E}=(\mathsf{KeyGen}, \mathsf{Enc}, \mathsf{Dec},
\mathsf{Eval})$ is a homomorphic encryption scheme with security parameter $\lambda$. Suppose
$\mathcal{A}$ is an adversary in the IND-CPA security game. The advantage of $\mathcal{A}$ with
respect to $\mathcal{E}$, denoted $Adv^{\mathcal{E}}_{\mathcal{A}}(\lambda)$, is defined to be:
\begin{align*}
Adv^{\mathcal{E}}_{\mathcal{A}}(\lambda) \overset{def}{=} \bigg| 2  \cdot \text{Pr}\bigg[  \text{IND-CPA}^{\mathcal{A}}_{\mathcal{E}} (1^{\lambda}) = 1 \bigg] - 1 \bigg|.
\end{align*}
\end{definition}

\begin{definition}[IND-CPA security]
\label{def:semsec}
Suppose $\mathcal{E}$, $\mathcal{A}$ and $\lambda$ are as in Definition~\ref{def:adv_INDCPA}.
$\mathcal{E}$ is \emph{IND-CPA secure} if the advantage $Adv^{\mathcal{E}}_{\mathcal{A}}(\lambda)$ for $\mathcal{A}$ in the IND-CPA security game is negligible.
\end{definition}

%
%

\section{A novel privacy-preserving protocol}
\label{sec:contrib}

We now describe the privacy-preserving biometric matching protocol. In fact we give two
descriptions: in \S\ref{subsec:informal} we give an informal introduction, explaining the
motivation for the design, and then in \S\ref{subsec:protocol} we give a formal description which
we use as the basis for the analysis in Section~\ref{sec:security}. For simplicity of presentation
we suppose that the public key $pk$ and the evaluation key $evk$ are equal.

\subsection{Informal description of the protocol}
\label{subsec:informal}

We describe a protocol involving two parties, a client $C$  and a server $S$, where $C$ is acting on behalf of user $U$\@.
$C$ wishes to access a certain service, not offered by $S$, which requires an initial authentication of the user $U$ associated with $C$ to $S$.
The process of authentication uses sensitive biometric data such as face images or iris information for $U$ that is gathered by $C$\@.
If $S$ successfully authenticates $U$, $S$ sends an ID token $\tau$, to $C$\@.
$C$ can now use $\tau$ to access the requested service.

Note that $C$ is trusted by $S$ to correctly gather a fresh biometric sample from $U$\@. In the
protocol, $S$ verifies that the gathered sample matches the appropriate user template, and also
authenticates $C$ to $S$. Note that the protocol neither provides authentication of $S$ to $C$ nor
provides encryption of transferred messages; it is implicitly assumed that these properties are
provided by the communications channel, e.g.\ using a server-authenticated TLS session.

In the description below, Step~0 (registration) is performed once before use of the protocol.
Steps~1-4 of the protocol are performed every time the user $U$ wishes to be authenticated to $S$ (via $C$).

\emph{Step~0: Registration}

$C$ generates a key pair $(sk_C, pk_C)$ for a homomorphic encryption scheme $\mathcal{E}$, and obtains by some means a biometric template $t$ for its associated user $U$\@.
$C$ then encrypts $t $ as $ \overline{t} \leftarrow  \mathsf{Enc}(pk_C, t)$ and sends $\overline{t}$ to $S$ via a trusted channel.
$S$ stores $\overline{t}$, and subsequently uses it for biometric matching when the protocol is executed (see Step~2).
In the remainder of this description we suppose that $S$, by some means, is assured of the identity of $U$ and that the encrypted biometric template $\overline{t}$ for $U$ is genuine.

\emph{Step~1: Initialisation}

$C$ takes a fresh biometric sample $s$ from $U$ and, using $\mathcal{E}$, computes an encrypted version $\overline{s} \leftarrow \mathsf{Enc}(pk_C, s)$ and sends it to $S$.

\emph{Step~2: Construction of the Matching Token} 

\quad \emph{Phase~1: Matching Computation} \quad

We suppose that $S$ is equipped with a biometric matching function $f \colon \mathcal{M} \times \mathcal{M} \rightarrow \mathbb{B}$ which inputs a biometric template and a biometric sample and outputs an indication of whether there is a sufficiently close match between them.
Suppose $\pi_f  \in \Pi_f$, where $\Pi_f$ 	is the circuit family associated with $\mathcal{E}$ which implements $f.$
$S$ now computes
\[ \overline{b} \leftarrow  \mathsf{Eval} (pk_C, \pi_f, \langle \overline{s}, \overline{t} \rangle),\]
where $\overline{b}$ is the encrypted version of a boolean $b$ indicating the success or not of the biometric matching, i.e.\ $\overline{b} \leftarrow \mathsf{Enc}(pk_C, f(s,t)).$

In a na\"{i}ve version of the protocol, $S$ now sends $C$ the encrypted matching result
$\overline{b}$; $C$ decrypts it to obtain $b \leftarrow \mathsf{Dec}(sk_C, \overline{b})$, and
sends $b$ to $S$. $S$ can now use $b$ to decide whether not to generate the ID Token $\tau$. For
obvious reasons this is not secure ($b$ is not authenticated), and hence we need a slightly more
elaborate protocol.

In order to enable $S$ to authenticate $C$, we introduce the notion of a \emph{Matching Token},
denoted by $y$.  In Phase~2 this token is constructed by $S$ as a function of $b$ (whilst still
encrypted) in such a way that $S$ can, when provided by $C$ with a decrypted version of the token
in Step~4, (a) verify its authenticity, and (b) determine the value of $b.$ \vspace{\baselineskip}

\quad \emph{Phase~2: Signature Computation} \quad
We suppose $S$ has an implementation of the function
\[ g(b, r_0, r_1) = (1 - b) \cdot r_0 + b \cdot r_1 .\]
$S$ first selects two random numbers $r_0 \overset{\mathdollar}{\leftarrow} \mathbb{B}^{\lambda}$ and $r_1 \overset{\mathdollar}{\leftarrow}
\mathbb{B}^{\lambda}$, and stores them for use in Step~4. $S$ next computes
\[\overline{r_0} \leftarrow \mathsf{Enc}(pk_C, r_0) ~~\mbox{and}~~ \overline{r_1} \leftarrow \mathsf{Enc}(pk_C, r_1).\]

In the encrypted domain of $ \mathcal{E}$ (under $pk_C$), $S$ now uses $\overline{b}\text{, }\overline{r_0} \text{ and } \overline{r_1}$ to compute the encrypted matching token $\overline{y}$ as:

\[ \overline{y} \leftarrow \mathsf{Eval}(pk_C, \pi_g, \langle \overline{b}, \overline{r_0}, \overline{r_1} \rangle) ,\]
where $\pi_g \in \Pi_g,$ the circuit family associated with $\mathcal{E}$ which implements $g.$
That is, $S$ obtains $\overline{y} \leftarrow \mathsf{Enc}(pk_C, g(b, r_0, r_1))$ although, of course, $S$ does not have access to $b$; i.e.\ at this stage $S$ does not know whether or not the biometric matching succeeded.
$S$ sends now $\overline{y}$ to $C$.

Note that this part of the protocol requires $S$ to retain the random values $r_0$ and $r_1$ until Step~4, and hence the protocol is stateful.

\emph{Step~3: Decryption of $\overline{y}$}

$C$ receives $\overline{y}$ from $S$ and computes
\[ y \leftarrow \mathsf{Dec}(sk_C, \overline{y}) .\]
At this point it is still the case that neither $C$ nor $S$ know whether the biometric matching succeeded.
$C$ only possesses a string which looks random, and $S$ cannot decrypt any data encrypted with $pk_C.$
$C$ now sends $y$ to $S.$

\emph{Step~4: Authentication of $C$}

\quad \emph{Phase~1: Verification} \quad

$S$ receives $y$ from $C$, and checks whether it is equal to $r_0$ or $r_1$.  If so, $S$ has
successfully authenticated $C$; if not $S$ rejects $C$.

\quad \emph{Phase~2: Token generation} \quad

$S$ generates an ID Token $\tau$ where $\tau \leftarrow \mbox{ACCEPT if } y = r_1 $, and $\tau
\leftarrow \mbox{REJECT}$ otherwise, and sends it to $C$. As a result, $C$ has a valid ID Token,
which can be used to access the desired service, if and only if the biometric matching was
successful and $S$ has authenticated $C$.

\subsection{Formal description of the protocol}
\label{subsec:protocol}

\begin{figure}[htb]
\frame{\includegraphics[width=\linewidth]{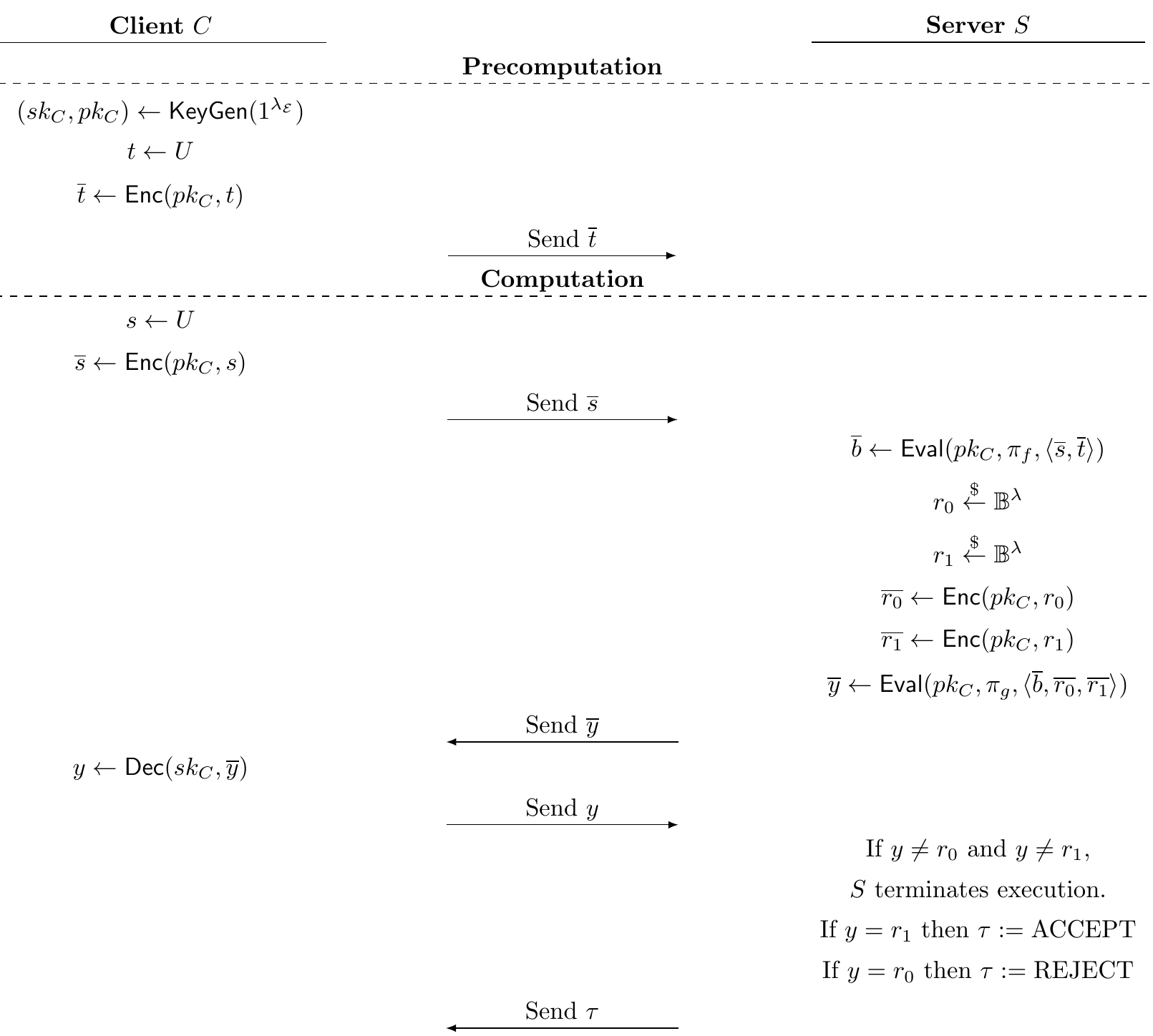}}
\caption{Protocol summary}
\label{fig:design1}
\end{figure}

We now formally present the protocol, referred to as $\mathcal{P}$. The protocol is summarised in
Figure~\ref{fig:design1}, where $\lambda_{\mathcal{E}}$ is the security parameter of $\mathcal{E}$.
Protocol initialisation, described immediately below, assumes Step~0 has been successfully
completed.

\emph{Input to $C$:}

$C$ has a biometric sample $s$, and a key pair $(sk_C, pk_C)$ generated with a homomorphic encryption scheme $\mathcal{E}$.
This is represented by the tuple $(s, sk_C)$.
We denote the plaintext space and the ciphertext space associated with $\mathcal{E}$ by $\mathcal{M}_{\mathcal{E}}$ and $\overline{\mathcal{M}_{\mathcal{E}}}$ respectively.

\emph{Input to $S$:}

$S$ has an encrypted biometric template $\overline{t} \leftarrow  \mathsf{Enc}(pk_C, t )$ generated by $C$ in a pre-computation phase.
This is represented by the tuple $(\overline{t}).$

The following functions are used by $S.$
\begin{itemize}
\item $f \colon \mathcal{M} \times \mathcal{M} \longrightarrow \mathbb{B}$ indicates whether or
    not two biometric values match, where $\mathcal{M}$ is the set of possible biometric values
    and an output of 1 indicates a match.
\item $g \colon \mathbb{B} \times \mathbb{B}^{\lambda} \times \mathbb{B}^{\lambda}
    \longrightarrow \mathbb{B}^{\lambda} $ creates a matching token $y$ from a boolean $b$ and
    two random numbers, where
\begin{align*}
g \colon (b, r_i, r_j)  & \longmapsto (1 - b) \cdot r_i + b \cdot r_j,~~\text{where}~ i,j \in \mathbb{N}.
\end{align*}
\end{itemize}

The above two initialisations are expressed formally as $\mathcal{P}: C(s) \leftrightarrow S(\overline{t})$\@.

\emph{Common input:}

Both parties know the homomorphic encryption scheme $\mathcal{E}$ and the public key $pk_C$ generated by $C$.

\emph{Protocol transcript:}

\label{para:transcript}
\begin{enumerate}[label=(\roman*)]
\item {[$C$ Pre-computation]:}
\begin{enumerate}[label=\alph*)]
\item $(sk_C, pk_C) \leftarrow \mathsf{KeyGen}(1^{\lambda_{\mathcal{E}}})$;
\item Take a fresh biometric sample $t$ from $U$ to be used as template;
\item  $\overline{t} \leftarrow \mathsf{Enc}(pk_C, t)$.
\end{enumerate}
\item {[$C$ $\longrightarrow$ $S$ Pre-computation]:}
\begin{enumerate}[label=\alph*)]
\item Send $\overline{t}$ to $S$.
\end{enumerate}
\end{enumerate}
\begin{enumerate}
\item {[$C$ $\longrightarrow$ $S$] } $C$ executes the following:
\begin{enumerate}
\item Take a fresh biometric sample $s$ from $U$;
\item Compute $\overline{s} \leftarrow \mathsf{Enc}(pk_C, s)$; 
\item Send $\overline{s}$ to $S$.
\end{enumerate}
\item {[$S$ $\longrightarrow$ $C$] } $S$ executes the following:
\begin{enumerate}
\item (\emph{Phase~1}) Compute $\overline{b} \leftarrow  \mathsf{Eval} (pk_C, \pi_f, \langle \overline{s}, \overline{t} \rangle )$; 
\item (\emph{Phase~2}) Generate $r_0, r_1 \overset{\mathdollar}{\leftarrow} \mathbb{B}^n$;
\item Compute $\overline{r_0} \leftarrow \mathsf{Enc}(pk_C, r_0)$;
\item Compute $\overline{r_1} \leftarrow \mathsf{Enc}(pk_C, r_1)$;
\item Compute $ \overline{y} \leftarrow \mathsf{Eval}(pk_C, \pi_g, \langle \overline{b}, \overline{r_0}, \overline{r_1} \rangle)$;
\item Send $\overline{y}$ to $C$.
\end{enumerate}
\item {[$C$ $\longrightarrow$ $S$] } $C$ executes the following:
\begin{enumerate}
\item Compute $ y \leftarrow \mathsf{Dec}(sk_C, \overline{y})$;
\item Send $y$ to $S$.
\end{enumerate}
\item {[$S$ $\longrightarrow$ $C$] } $S$ executes the following:
\begin{enumerate}
\item If $y\not=r_0$ and $y\not=r_1$, $S$ terminates execution;
\item Compute $\tau \leftarrow
\begin{cases} \text{ ACCEPT }  & \text{if } y = r_1 ,\\
  \text{ REJECT } & \text{if } y=r_0;
\end{cases}$
\item Send $\tau$ to $C$.
\end{enumerate}
\end{enumerate}

\subsection{Proof of knowledge}
\label{subsec:proof_of_knowledge}

The protocol $\mathcal{P}$ is an instance of an \emph{interactive proof system}, as defined by
Menezes et al. \cite[Chapter~10]{DBLP:books/crc/MenezesOV96}. We next show that $\mathcal{P}$ is a
\emph{proof of knowledge}, i.e.\ it has the properties of \emph{completeness} and \emph{soundness}.
The following definition is adapted from \cite[Chapter~10]{DBLP:books/crc/MenezesOV96}.

\begin{definition}[Completeness]
\label{def:completeness} An authentication protocol is \emph{complete} if, given an honest client
$C$ and an honest server $S$, the protocol succeeds with overwhelming probability (i.e.\ $S$
accepts $C$'s claim).
\end{definition}

\begin{theorem}[Completeness]
\label{thr:completeness}
The protocol $\mathcal{P}$ is \emph{complete}.
\end{theorem}

\begin{IEEEproof}
Suppose $\mathcal{P}$ is run with an honest client $C$ that sends a validly constructed value
$\overline{s} \leftarrow \mathsf{Enc}(pk_C, s)$ for a sample $s$ to server $S.$ We consider two
cases.
\begin{enumerate}[label=(\alph*)]
\item Suppose the sample $s$ matches the template $t$, i.e.\ suppose $f(s,t)=1.$
Then, by definition, $\overline{b} = \mathsf{Enc}(pk_C,1)$, and thus $\overline{y}=\overline{r_1}.$
Hence, if $y \leftarrow \mathsf{Dec}(sk_C, \overline{y})$ then $y=r_1.$
Thus, $S$ accepts $C$.
\item Suppose the sample $s$ does not match the template $t$, i.e.\ suppose $f(s,t)=0$\@.

Then, by definition, $\overline{b} = \mathsf{Enc}(pk_C,0)$, and thus $\overline{y}=\overline{r_0}$\@.
Hence, if $y \leftarrow \mathsf{Dec}(sk_C, \overline{y})$ then $y=r_0$\@.
Thus, $S$ does not accept $C.$
\end{enumerate}
That is, $S$ accepts $C$ if and only if the sample $s$ matches the template $t$. 
\end{IEEEproof}

The following definition is adapted from \cite[Chapter~10]{DBLP:books/crc/MenezesOV96}.

\begin{definition}[Soundness]
\label{def:soundness} An authentication protocol is \emph{sound} if there exists an expected
polynomial time algorithm $\mathcal{A}$ with the following property: if a dishonest client $C'$
(impersonating $C$) can with non-negligible probability successfully execute the protocol with $S$,
then $\mathcal{A}$ can be used to extract from $C'$ knowledge (essentially equivalent to $C$'s
secret) which with non-negligible probability allows successful subsequent protocol executions.
\end{definition}

We first need the following preliminary result.

\begin{lemma} \label{lem:acceptance}
Suppose a client $C^*$ engages in the protocol $\mathcal{P}$ with the server $S$, using sample
$s_{C^*}$, and that $S$ accepts $C^*$.  It follows that:
\begin{enumerate}[label=(\alph*)]
\item the sample $s_{C^*}$ matches the template $t$ held by $S$;
\item $C^*$ has access to the value $r_1$ chosen by $S$ in Step 2b of $\mathcal{P}$.
\end{enumerate}
\end{lemma}

\begin{IEEEproof}
\begin{enumerate}[label=(\alph*)]
\item Since $S$ accepts $C^*$, it immediately follows from Theorem~\ref{thr:completeness} that
    the sample $s_{C^*}$ matches the template $t$.
\item In Step 4 of $\mathcal{P}$, $S$ accepts $C^*$ if and only if the value $y$ sent by $C^*$
    to $S$ in Step 3 equals $r_1$. The result follows. 
\end{enumerate}
\end{IEEEproof}

We can now give our main result.

\begin{theorem}
\label{thr:soundness}
The protocol $\mathcal{P}$ is \emph{sound}.
\end{theorem}

\begin{IEEEproof}
Suppose $\mathcal{P}$ is run with a dishonest client $C'$, impersonating an honest client $C$, that
sends a validly constructed value $\overline{s_{C'}} \leftarrow \mathsf{Enc}(pk_{C}, s_{C'})$ for a
sample $s_{C'}$ to server $S$ in Step 1 of $\mathcal{P}$.  Suppose also that there is a
non-negligible probability that $C'$ is accepted. We need to establish that $C'$ can, with
non-negligible probability, engage in further successful protocol executions with $S$.

Since $S$ accepts $C'$ in the protocol execution with non-negligible probability, by
Lemma~\ref{lem:acceptance} we know that $C'$ with non-negligible probability has access to $r_1$,
which was provided to $C'$ in encrypted form in Step 2 of $\mathcal{P}$. Hence $C'$ must have
access to an oracle $\mathcal{O}$ that, given an input encrypted using $C$'s public key, with
non-negligible probability returns its decrypted version.

Assume a subsequent instance of the same protocol $\mathcal{P}$.
\begin{enumerate}
\item In Step~1, $C'$ uses the sample $s_{C'}^* = s_{C'}$, computes $\overline{s_{C'}^*}$ using
    the public key of $C$, and sends it to $S$.
\item Step~2 is executed as specified by $S$, where the two random values chosen by $S$ are
    denoted by $r_0^*$ and $r_1^*$.  Clearly $s_{C'}^*$ matches $t$ (from (a) above), and hence
    the value $\overline{y^*}$ sent to $C'$ will satisfy $y^*=r^*_1$.
\item In Step~3, $C'$ uses oracle $\mathcal{O}$ which will, with non-negligible probability,
    correctly decrypt $\overline{y^*}$; that is, the value $y^*$ output by $\mathcal{O}$ will
    satisfy $y^*=r^*_1$ with non-negligible probability. $C'$ then sends $y^*$ to $S$.
\item In Step~4, since $y^*=r^*_1$ with non-negligible probability, $S$ will accept $C'$ with
    non-negligible probability.
\end{enumerate}
That is, there exists a PPT algorithm $\mathcal{A}$, using $\mathcal{O}$ as a subroutine, that for
any instance of $\mathcal{P}$ can be used to arrange that $C'$ will be accepted by $S$ with
non-negligible probability. 
\end{IEEEproof}

\section{Security properties}
\label{sec:security}

\subsection{Security model}
\label{subsec:model}

%

%
%
We suppose the protocol $\mathcal{P}$ is carried out in the \emph{real world} between a challenger and an adversary.
In the real world, adversaries can play the role of the client or the server.
We suppose adversaries are \emph{static}, i.e.\ they cannot change their role within an instance of the protocol, and cannot play both roles at the same time.
We distinguish between two classes of adversary:
\begin{itemize}
\item \emph{Honest-but-curious adversaries} execute a protocol honestly, although `on the side' they can make any other calculations with the purpose of obtaining information to which they are not entitled.
\item \emph{Malicious adversaries} execute a protocol in ways not permitted in the
    specification, perform any calculations, and use any means to obtain information.
\end{itemize}
In our setting, we model an honest-but-curious adversary.

\subsection{Privacy of the biometric data}
\label{subsec:privacy}
One of the main goals of $\mathcal{P} $ is to give $C$ (and $U$) assurance regarding the privacy of biometric data shared with $S$, i.e.\ all samples and templates.
As we next show, this property relies on the IND-CPA security (see Definition~\ref{def:semsec}) of the homomorphic encryption scheme.

\begin{definition}[Privacy-preserving]
\label{def:priv-pres} If a biometric authentication protocol preserves the privacy of the biometric
data of the client against a honest-but-curious adversary (a server or external party), then the protocol is \emph{privacy-preserving}.
\end{definition}

\begin{definition}[Privacy-preserving game]
\label{def:priv-pres-game} Suppose the precomputation phase of the protocol $\mathcal{P}$ is run
with an honest client $C$ that sends a validly constructed encrypted value $\overline{s} \leftarrow
\mathsf{Enc}(pk_C, s)$ for template $\overline{t} \leftarrow \mathsf{Enc}(pk_C, t)$ to server $S$.
Suppose also that $\mathcal{A}$ is a PPT adversary. The \emph{privacy-preserving game} is as
follows.
\begin{enumerate}
\item A challenger chooses $i \in \mathbb{B}$ uniformly at random and generates two distinct
    samples $\{ s_0, s_1 \}$ as follows.
\begin{enumerate}[label=(\alph*)]
\item $f(s_i,t)=1$, and
\item $f(s_{1-i},t)=0.$
\end{enumerate}
\item The challenger encrypts the two samples as $\overline{s_0} \leftarrow \mathsf{Enc}(pk_C, s_0)$ and $\overline{s_1} \leftarrow \mathsf{Enc}(pk_C, s_1).$
\item The challenger sends $\{ \overline{s_0}, \overline{s_1}, \overline{t} \}$ to $\mathcal{A}.$
\item $\mathcal{A}$ outputs a pair $(\overline{s_j}, \overline{t})$, where $j \in \mathbb{B},$ and wins the game if $i=j.$
\end{enumerate}
We denote this security game by $\text{PRI-PRE}^{\mathcal{A}}_{\mathcal{P}} (1^{\lambda})$, where $\lambda$ is the security parameter of the homomorphic encryption scheme $\mathcal{E}$ and a win in an instance of this security game by $\text{PRI-PRE}^{\mathcal{A}}_{\mathcal{P}} (1^{\lambda}) = 1$.
\end{definition}

\begin{definition}[Advantage for the PRI-PRE game]
\label{def:priv-pres-adv} Suppose that $\mathcal{P}$, $\lambda$ and $\mathcal{A}$ are as in
Definition~\ref{def:priv-pres-game}. The \emph{advantage of the adversary with respect to}
$\mathcal{P}$, denoted by $Adv^{\mathcal{P}}_{\mathcal{A}}(\lambda)$, is defined to be:
\begin{align*}
Adv^{\mathcal{P}}_{\mathcal{A}}(\lambda) \overset{def}{=} \bigg| 2  \cdot \text{Pr}\bigg[  \text{PRI-PRE}^{\mathcal{A}}_{\mathcal{P}} (1^{\lambda}) = 1 \bigg] - 1 \bigg|.
\end{align*}
\end{definition}

\begin{definition}[PRI-PRE security]
\label{def:pri-pre} Suppose $\mathcal{P}$, $\mathcal{A}$ and $\lambda$ are as in
Definition~\ref{def:priv-pres-adv}. If the advantage $Adv^{\mathcal{P}}_{\mathcal{A}}(\lambda)$ for
$\mathcal{A}$ in the PRI-PRE game is negligible, then $\mathcal{P}$ is \emph{PRI-PRE}, i.e.\
privacy-preserving.
\end{definition}

\begin{theorem}
\label{thr:priv-pres}
The protocol $\mathcal{P}$ is privacy-preserving.
\end{theorem}

\begin{IEEEproof}
Suppose that the protocol $P$ is not privacy-preserving, i.e.\ by Definition~\ref{def:pri-pre},
there exists an adversary $\mathcal{A}$ that has a non-negligible advantage in the
privacy-preserving game. By definition this means that $\mathcal{A}$ has a distinguisher
$\mathcal{D}$ that distinguishes, with non-negligible probability, which of two encrypted samples
$\overline{s_0}$ and $\overline{s_1}$ will match an encrypted template $\overline{t}$.

We next construct an adversary $\mathcal{B}$ against the IND-CPA security of $\mathcal{E}$\@.
Suppose $\mathcal{B}$ generates a triple of values $(s,s',t)$ satisfying $f(s,t)=1$ and
$f(s',t)=0$. $\mathcal{B}$ now submits the pair $(s,s')$ to a challenger in the IND-CPA security
game. $\mathcal{B}$ receives back from the challenger the ciphertext $\overline{s^*}$, where
$s^*$ equals either $s$ or $s'$ (with equal probability).

$\mathcal{B}$ first computes $\overline{s}$ and $\overline{t}$ from $s$ and $t$, and then runs the
distinguisher $\mathcal{D}$ with inputs $\overline{s^*}$ and $\overline{s}$ as the encrypted
samples and $\overline{t}$ as the encrypted template. If $\mathcal{D}$ returns $\overline{s^*}$
(which we call event $e_X$), then $\mathcal{B}$ outputs $(s,\overline{s^*})$ in the IND-CPA game.
If $\mathcal{D}$ returns $\overline{s}$ (which we call event $e_Y$), then $\mathcal{B}$ outputs
$(s',\overline{s^*})$ in the IND-CPA game.

To evaluate the probability that $\mathcal{B}$ wins the game, we consider two cases.
\begin{itemize}
\item Suppose $s^*=s$ (event $e_A$ which has probability 0.5). Then the two encrypted samples
    $\overline{s^*}$ and $\overline{s}$ submitted to $\mathcal{D}$ both match the template.
    Hence the probability that $\mathcal{D}$ will return $\overline{s^*}$ (event $e_X$) = the
    probability it returns $\overline{s}$ (event $e_Y$) = 0.5.
\item Suppose $s^*=s'$ (event $e_B$ which also has probability 0.5). Then of the two encrypted
    samples $\overline{s^*}$ and $\overline{s}$ submitted to $\mathcal{D}$, only $\overline{s}$
    will match the template. Hence the probability that $\mathcal{D}$ will return
    $\overline{s}$ (event $e_Y$) is $0.5+p$, where $p>0$ is non-negligible (this follows since
    $\mathcal{D}$ is a distinguisher).
\end{itemize}

Hence we have:
\begin{eqnarray*}
\mbox{Pr}(e_A\wedge e_X) = \mbox{Pr}(e_A)\mbox{Pr}(e_X) = 0.5^2 = 0.25;~~~\mbox{and}\\
\mbox{Pr}(e_B\wedge e_Y) = \mbox{Pr}(e_B)\mbox{Pr}(e_Y) = 0.5(0.5+p) = 0.25+0.5p.
\end{eqnarray*}

If $e_X$ occurs then, by assumption, $\mathcal{B}$ outputs $(s,\overline{s^*})$ in the IND-CPA
game. The probability this wins is simply $\mbox{Pr}(e_A|e_X)$.
Similarly, if $e_Y$ occurs then the probability of $\mathcal{B}$ winning is $\mbox{Pr}(e_B|e_Y)$.
Hence, since events $e_X$ and $e_Y$ are mutually exclusive, the probability that $\mathcal{B}$ wins
the game is:
\[ \mbox{Pr}(e_A|e_X)\mbox{Pr}(e_X) + \mbox{Pr}(e_B|e_Y)\mbox{Pr}(e_Y) \]
\[ = \mbox{Pr}(e_A\wedge e_X)+\mbox{Pr}(e_B\wedge
e_Y) = 0.5(1+p). \]

By definition the advantage for $\mathcal{B}$ is $2(0.5(1+p))-1=2p$, which is non-negligible since
$p$ is non-negligible. This contradicts the assumption that $\mathcal{E}$ is IND-CPA secure, and
hence $\mathcal{P}$ is privacy-preserving. 
\end{IEEEproof}

\subsection{Entity authentication}
\label{subsec:entity-auth}

We next show that Steps~2--4(a) of $\mathcal{P}$ constitute a secure authentication protocol. We follow the approach of Boyd
et al.\ \cite{DBLP:series/isc/BoydMS20}, based on the Bellare-Rogaway model
\cite{DBLP:conf/crypto/BellareR93}, adapting a proof of Blake-Wilson and Menezes
\cite{DBLP:conf/spw/Blake-WilsonM97}. We first give an informal definition of entity
authentication.

\begin{definition}[Menezes et al.\ \cite{DBLP:books/crc/MenezesOV96}]
\label{ent-auth-gen}
\emph{Entity authentication} is the process whereby one party is assured (through acquisition of corroborative evidence) of the identity of a second party involved in a protocol, and that the second has actually participated (i.e.\ is active at, or immediately prior to, the time the evidence is acquired).
\end{definition}

Steps~2--4(a) of $\mathcal{P}$ by design constitute a unilateral entity authentication protocol, i.e.\ only $C$
authenticates to $S$\@.
Before formally defining the authentication notion, we need the concept of
matching conversations due to Bellare and Rogaway \cite{DBLP:conf/crypto/BellareR93}. We suppose
that an adversary $\mathcal{A}$ has access to an infinite family of oracles denoted by
$\Omega^i_{a,b}$, where $a$ and $b$ are in the space of participants of a protocol, $i \in
\mathbb{N}$ denotes the $i$-th instance of a protocol, and the oracle behaves as if entity $a$ is performing protocol $\mathcal{P}$ in the belief it is communicating with the entity $b$ for $i$th time.

\begin{definition}[Conversation]
\label{def:conv}
For any oracle $\Omega^i_{a,b}$, its \emph{conversation} for instance $i$ is the following $n$-tuple
\[ K = (t_1, \alpha_1, \beta_1), (t_2, \alpha_2, \beta_2), ..., (t_n, \alpha_n, \beta_n)\]
where at time $t_j$, the oracle $\Omega^i_{a,b}$ received $\alpha_j$ and sent $\beta_j$ $(1 \leq j \leq n)$\@.
\end{definition}

We can now define matching conversations, again following Bellare and Rogaway \cite[Definition~4.1]{DBLP:conf/crypto/BellareR93}.
We assume that the number of moves $n$ in a protocol is odd ($n$ even is investigated by Boyd et al. \cite{DBLP:series/isc/BoydMS20}).

\begin{definition}[Matching conversations]
\label{def:matching_conv} Suppose $P$ is a $n$-move protocol, where $n = 2k-1$ for some integer
$k$\@. Run $P$ and suppose oracles $\Omega^i_{a,b}$ and $\Omega^j_{b,a}$ engage in conversations
$K_i$ and $K_j$, respectively.
If there exist $ t_0 < t_1 < ... < t_n$ and $\alpha_1, \beta_1, ..., \alpha_k, \beta_k$
such that \ $K_i$ is prefixed by

\[ (t_0, \emptyset, \alpha_1), (t_2, \beta_1, \alpha_2), (t_4, \beta_2, \alpha_3), ...,  (t_{2k-2}, \beta_{k-1}, \alpha_k)\]
and $K_j$ is prefixed by
\[t_1, \alpha_1, \beta_1), (t_3, \alpha_2, \beta_2), (t_5, \alpha_3, \beta_3), ...,  \] 
\[ (t_{2k-3}, \alpha_{k-1}, \beta_{k-1}) , (t_{2k-1}, \alpha_{k}, *) \] 
then $K_j$ is a \emph{matching conversation} to $K_i$\@.



$\emptyset$ means that the oracle has no input, because it initiates the protocol; we call it an
\emph{initiator oracle}; otherwise, an oracle is a \emph{responder oracle}. $*$ means that the
oracle has no output, because the protocol ends with this last move.
\end{definition}

Informally, this means that conversation $K_j$ of $\Omega^j_{b,a}$ (a responder oracle) matches conversation $K_i$ of $\Omega^i_{a,b}$ (an initiator oracle). 
We also need the following definition, which has been modified for the unilateral (as opposed to mutual) authentication case.

\begin{definition}[No match]
\label{def:no_match}
Suppose $P$ is a protocol and $\mathcal{A}$ is an adversary.
Suppose also that when $P$ is run against $\mathcal{A}$ there exists an initiator oracle $\Omega^i_{a,b}$ with a conversation $K_i$ in the ACCEPT state but no oracle $\Omega^j_{b,a}$ has a conversation matching with $K_i$\@.
We denote this event by $\texttt{No-Match}^{\mathcal{A}}_P$ and its probability by $\text{Pr}(\texttt{No-Match}^{\mathcal{A}}_P)$\@.
\end{definition}

These preliminaries enable us to state the following key definition.
Note that this definition corresponds to the case where the protocol responder (entity $b$) is authenticated by the protocol initiator (entity $a$), i.e.\ in the case of protocol $\mathcal{P}$ where the server is entity $a$ and the client is entity $b$\@. 

\begin{definition}[Secure unilateral authentication protocol]
\label{def:sec_auth_conv} A protocol $P$ is a secure unilateral entity authentication protocol if
for every adversary $\mathcal{A}$:
\begin{enumerate}
\item If $\Omega^{i}_{a, b}$ and $\Omega^{j}_{b, a}$ have matching conversations, then the
    initiator oracle $\Omega^{i}_{a, b}$ accepts;
\item $\text{Pr}(\texttt{No-Match}^{\mathcal{A}}_P)$ is negligible.
\end{enumerate}
\end{definition}

The first condition refers to completeness. The second condition says that the only way for an
adversary to corrupt an honest responder oracle to the ACCEPT state is to relay the messages in the
protocol without modification, i.e.\ an adversary can only observe and relay messages.

We can now state the main result.

\begin{theorem}
\label{thr:red_nomatch_aut} If $\mathcal{E}$ is IND-CPA, then Steps~2--4(a) of $\mathcal{P}$ form a
secure unilateral authentication protocol.
\end{theorem}

\begin{IEEEproof}
Since for the purposes of the Theorem we are ignoring Steps~1, 4(b) and 4(c) of $\mathcal{P}$, the server is the protocol initiator and the client is the responder, although the reverse is true for $\mathcal{P}$ in its entirety.
Suppose $\lambda$ is the security parameter of the underlying homomorphic encryption scheme $\mathcal{E}$.
Suppose also that Steps~2--4(a) of $\mathcal{P}$ do not form a secure authentication protocol.
From Theorem~\ref{thr:completeness}, we know that $\mathcal{P}$ is complete, i.e.\ that the first
condition of Definition~\ref{def:sec_auth_conv} holds. Thus the second condition does not hold,
i.e.\ there exists a PPT adversary $\mathcal{A}$ such that
$\text{Pr}(\texttt{No-Match}^{\mathcal{A}}_{\mathcal{P}})$ is non-negligible.

We say that $\mathcal{A}$ \emph{succeeds} against $\Omega^i_{a,b}$ if, at the end of $\mathcal{A}$'s operation, there exists an initiator oracle $\Omega^i_{a,b}$ with a conversation $K_i$ in the ACCEPT state but no oracle $\Omega^j_{b,a}$ has a conversation $K_j$ matching with $K_i$\@.
We denote the probability that $\mathcal{A}$ succeeds against the initiator oracle $\Omega^i_{a,b}$ by $\text{Pr}(\mathcal{A} \text{ succeeds}) = p$\@.
Then, by assumption, $p$ is non-negligible. 
Suppose also $\mathcal{A}$ possesses the public key $pk_{\mathcal{A}}$ of a genuine client.
We next construct an adversary $\mathcal{B}$ from $\mathcal{A}$ against the IND-CPA security of $\mathcal{E}$\@.

We consider the details of the conversation of the oracle $\Omega^i_{S,C}$. Since we only consider
Steps~2--4(a) of $\mathcal{P}$, we have $n=3$.  Suppose the conversation for $\Omega^i_{S,C}$ is
\[ K = (t_0, \emptyset, \alpha_1), (t_2, \beta_1, \alpha_2) \]
where at time $t_0$, the oracle sent $\alpha_1$ and at time $t_2$ the oracle received $\beta_1$ and sent $\alpha_2$\@.
Then it follows that we have $\alpha_1=\overline{y}=\mathsf{Enc}(pk_C,r_w)$ (where
$w$ is 0 or 1), $\beta_1=y$, and $\alpha_2=*$), where we ignore the ID token $\tau$ since its
construction is independent of the design of the protocol.

Since $\mathcal{A}$ is successful against $\Omega^i_{S,C}$ with probability $p$, it follows that
$y\in\{r_0,r_1\}$ with probability $p$.  Since $r_0$ and $r_1$ are chosen uniformly at random for
each conversation instance, and since we are also assuming that there is no matching conversation,
$\mathcal{A}$ must have a means for recovering $r_w$ from $\mathsf{Enc}(pk_C,r_w)$ which works
with probability at least $p$. Hence $\mathcal{A}$ must have access to an oracle $\mathcal{O}$
which, when given an input encrypted using the public key of $C$, with non-negligible probability
returns its decrypted version. However, since $\mathcal{A}$ does not have access to the private
key of $C$, this oracle can immediately be used to construct an adversary $\mathcal{B}$ against the
IND-CPA security of $\mathcal{E}$\@. This gives the desired contradiction and hence it follows that
$\mathcal{P}$ is a secure unilateral authentication protocol. 
\end{IEEEproof}

\section{Implementation}
\label{sec:implementation}

The protocol has been implemented using the C/C++ Fully Homomorphic Encryption over the Torus (TFHE) library due to Chillotti et al. \cite{TFHE}.
One feature of TFHE is that it implements \emph{gate bootstrapping}, i.e.\ at each evaluated gate
the bootstrapping method is executed. This enables the evaluation of arbitrary circuits on
encrypted data. In practice, TFHE offers the fastest gate bootstrapping in the literature, namely
of the order of 13 milliseconds per gate on a single core; however, ``bootstrapped bit operations
are still about one billion times slower than their plaintext equivalents''
\cite{DBLP:journals/joc/ChillottiGGI20}.

In Section~\ref{sec:prelim}, we described a homomorphic encryption scheme as a public key encryption system.
The TFHE scheme is symmetric but can easily be used in the context of $\mathcal{P}$ because it provides a pair of keys: a secret key $sk$ and a cloud key $ck$\@.
In the context of $\mathcal{P}$ (see Section~\ref{sec:contrib}), $sk$ is kept secret and used by the client $C$ to encrypt and decrypt data.
$C$ sends $ck$ to the server $S$ during the registration phase.
$S$ is then able to compute arbitrary circuits on data encrypted under $sk$ using $ck$ without being able to decrypt them.
For further information on the design and security of TFHE see Chillotti et al. \cite{DBLP:journals/joc/ChillottiGGI20}.

\subsection{Biometric matching}
\label{subsec:biometric_match}


We chose facial recognition as the biometric method for our proof-of-concept implementation for two main reasons: it is a mature technology (see, for example, the NIST report \cite{NIST_BM}) and one that suits the homomorphic setting.
For our purposes, facial samples and templates are vectors $\mathbf{x} = \langle x_1, ..., x_n \rangle \in (\mathbb{Z}_m)^n$, where $\mathbb{Z}_m$ is the set of the integers modulo $m$ (for some $m$).
Samples and templates are compared using Euclidean distance, as defined below.

\begin{definition}[Euclidean distance]
\label{def:euclidean_distance}
Suppose $\mathbf{x}, \mathbf{y} \in (\mathbb{Z}_m)^n$\@.
The \emph{Euclidean distance} between $\mathbf{x}$ and $\mathbf{y}$ is defined to be:
\[ \Delta_{\mathbf{x},\mathbf{y}} = \sqrt{\sum_{i=1}^n (y_i - x_i)^2} .\]
\end{definition}


To simplify calculations, we used the square of the distance as the metric. As in the following
definition, a sample and a template are deemed to match if the (square of) the distance is at most
$B$, for some $B$\@.

\begin{definition}[Match]
\label{def:match} A pair of vectors $\mathbf{x},\mathbf{y} \in (\mathbb{Z}_m)^n$ are said to
\emph{match} if and only if $(\Delta_{\mathbf{x},\mathbf{y}})^2 \leq \mathcal{B}$\@.
\end{definition}
The function $f$, defined in \S\ref{subsec:protocol}, is implemented in accordance with Definition~\ref{def:match} as follows:
$f \colon (\mathbb{Z}_m)^n \times (\mathbb{Z}_m)^n \longrightarrow \mathbb{B}$, where $f(\mathbf{x}, \mathbf{y})=1$ if and only if $(\Delta_{\mathbf{x},\mathbf{y}})^2 \leq \mathcal{B}$, and we assume this implementation throughout Section~\ref{sec:implementation}.
The algorithm used to implement $f$ is given in Appendix~\ref{apx:alg} (see Algorithm~\ref{alg:f}).
%
%
%
%

For comparison purposes, when verifying the correctness of the implementation, we also implemented the \emph{Manhattan distance}, defined below.

\begin{definition}[Manhattan distance]
\label{def:manhattan_dist} Suppose $\mathbf{x}, \mathbf{y} \in (\mathbb{Z}_m)^n$, and let $|z|$
denote the absolute value of $z$\@. The \emph{Manhattan distance} between $\mathbf{x}$ and
$\mathbf{y}$ is defined to be:
\[ \delta_{\mathbf{x},\mathbf{y}} = \sum_{i=1}^n |y_i - x_i|.\]
\end{definition}

%

\subsection{Results}
\label{subsec:results}

To obtain performance results, the implementation was run on an Ubuntu 20.04.1 LTS 64-bit machine
with 8 GB of RAM and a four-core Intel(R) Core(TM) i3-6100CPU @ 3.70GHz.
TFHE was used with the default parameter, which achieves 110-bit cryptographic security \cite{TFHE}.
We chose to use biometric vectors of length 128 (i.e.\ $n=128$) because it is a likely real-world value.

To obtain timing figures, we first measured the `homomorphic' (ciphertext domain) computation times
for most of the arithmetic and bit comparison subroutines given in Appendix~\ref{apx:alg}. For
comparison purposes, we also implemented and measured the performance of all the subroutines in the
plaintext domain. Table~\ref{tab:results} summarises the results.

It is clear that homomorphic computations have a substantial performance cost, with an order of
magnitude of at least $10^6$. This finding is in line with previous work
\cite{cryptoeprint:2015:1192}, despite the optimisations included in the TFHE library
\cite{DBLP:conf/asiacrypt/ChillottiGGI16}.

\begin{table}[htbp]
\caption{Performance results for basic operations} \label{tab:results}
\begin{center}
\begin{tabular}{|l|r|r|}
\hline
Subroutines            & \vtop{\hbox{\strut Execution time}\hbox{\strut on plaintexts}\hbox{\strut (in nanoseconds)}}& \vtop{\hbox{\strut Execution time}\hbox{\strut on ciphertexts}\hbox{\strut (in seconds)}} \\ \hline
$n$-bit addition       & 335                                              & 9                                        \\ \hline
Two's complement       & 422                                              & 10                                        \\ \hline
Absolute value         & 396                                              & 10                                        \\ \hline
$n$-bit subtraction    & 1108                                              & 30                                        \\ \hline
$n$-bit multiplication & 2094                                             & 206                                        \\ \hline
Manhattan distance     & 210370                                           	& 5049                                       \\ \hline
Euclidean distance     & 425022                                              & 33536                                     \\ \hline
\end{tabular}%
\end{center}
\end{table}

Building on the implementations of fundamental operations, we implemented a naive version of
$\mathcal{P}$. The performance results are shown in Table~\ref{tab:P_results}, and confirm that the
current proof-of-concept implementation is certainly not practical, and needs considerable
optimisation in order to be usable in practice. For comparison we also show computation results in
the plaintext domain. Note that none of the performance results given in Table~\ref{tab:results}
include the encryption and decryption time.

\begin{table}[htbp]
\caption{Performance results for the protocol $\mathcal{P}$ and its underlying functions}
\begin{center}
\begin{tabular}{|l|r|r|}
\hline
Subroutines            & \vtop{\hbox{\strut Execution time}\hbox{\strut on plaintexts}\hbox{\strut (in microseconds)}}& \vtop{\hbox{\strut Execution time}\hbox{\strut on ciphertexts}\hbox{\strut (in seconds)}} \\ \hline
Function $f$       		& 790                                              & 34308                                        \\ \hline
Function $g$       		& 5                                                & 456                                        \\ \hline
Protocol $\mathcal{P}$	& 810                                              & 34765                                    \\ \hline
\end{tabular}%
\label{tab:P_results}
\end{center}
\end{table}

These results demonstrate the importance of optimising the design of an algorithm and its
implementation. The performance results are not only due to the homomorphic paradigm, but also
because we implemented the most naive routines without any optimisations or parallelisations. We
project from those results that with an optimised and targeted implementation $\mathcal{P}$ could
be practical in the real world.

To conclude, we showed that, implemented naively, homomorphic encryption does not meet the
performance criteria for practical use, since a user cannot wait for a  few hours to be authenticated in
most (if not all) authentication use cases. Indeed, Nah \cite{DBLP:conf/amcis/Nah03} showed that a
typical user will not tolerate a wait of more than two seconds for a web page to appear.
Nonetheless, there are considerable possibilities for optimisation, and the implementation and
design of $\mathcal{P}$ can be enhanced in various ways, as we next briefly discuss.

\subsection{Possible optimisations}

The most obvious improvement would be from the \emph{algorithmics} perspective. As explained above
all the subroutines are implemented in a very na\"{i}ve way.

There exist various public libraries that could be used to add \emph{parallel computing} features.
One example would be a C++ library such as OpenMP. Many of the subroutines have \textsl{for} loops
in which all execution instances are independent.

Finally, perhaps the most effective optimisation would be to \emph{mix} the FHE schemes, as
proposed by Boura et al.\ \cite{cryptoeprint:2018:758,DBLP:journals/jmc/BouraGGJ20}. Existing
libraries are optimised for certain targeted homomorphic computations; the main idea is to switch
between libraries, choosing the most efficient for each homomorphic computation. In our case, the
arithmetic subroutines would be faster on libraries other then TFHE; however, bit comparisons are
much better handled by the TFHE library. This idea is practically effective, as shown by Lou et
al.\ \cite{DBLP:conf/nips/LouFF020} who present \textsf{Glyph}, a tool which switches between TFHE
\cite{TFHE} and BGV cryptosystems \cite{DBLP:conf/innovations/BrakerskiGV12}.

\section{Conclusions and future work}
\label{sec:conclu}

We presented the design and a proof-of-concept implementation of a novel privacy preserving
authentication protocol based on fully homomorphic encryption. Human authentication is based on
biometric matching, implemented in the proof-of-concept using face matching. In the implementation,
all underlying operations are executed using FHE, including biometric matching, Euclidean distance
computation, and integer comparison. We showed that the protocol is privacy-preserving and a secure
unilateral authentication protocol if the underlying homomorphic encryption scheme is IND-CPA.

The implementation results are for a naive and unoptimised version, i.e.\ the worst-case scenario.
However, producing it involved developing a set of elementary routines in the ciphertext domain
that can be used as low-level building blocks in other applications. The results confirm that FHE
is not practical in a naive worst-case model, and real-world implementations would require
optimisations. However, the results suggest that, with already identified improvements, the
protocol can be made ready for real-world adoption.

There are number of possible directions for future work in improving performance. First, as
identified in \S\ref{subsec:results}, mixing FHE schemes to take advantage of the best of each
scheme (see \cite{DBLP:journals/jmc/BouraGGJ20,DBLP:conf/nips/LouFF020}) would significantly
benefit performance without compromising the IND-CPA security of the homomorphic encryption scheme.
Better algorithmics and implementation design is also an obvious target for improvement. Another
possibility would be to change the biometric matching paradigm. Deep Learning is known to be useful
in this context, and the performance in particular for face matching has been much improved
recently thanks to initiatives such as that of NIST\footnote{See
\url{https://www.nist.gov/speech-testimony/facial-recognition-technology-frt-0} for more details.}.
However, when such deep learning techniques are used in combination with homomorphic encryption,
only the inference phase is run homomorphically and the training phase is run on clear data (see
e.g.\ \cite{DBLP:conf/crypto/BourseMMP18,DBLP:conf/icml/Gilad-BachrachD16}). To achieve the level
of security we showed in this paper with FHE, both phases need to be executed in the ciphertext
domain. However, encrypting both phases may not be straightforward to achieve, as recent experience
shows that it is costly \cite{DBLP:conf/nips/LouFF020,DBLP:conf/cvpr/NandakumarRPH19}, despite
improvements in making FHE practical.

\bibliographystyle{IEEEtran}
\bibliography{references_abbr}

\appendix

\section{Circuits and circuit families}
\label{apd:circuits}

We next formally introduce notions related to circuits.
For more complete versions of these definitions, see Vollmer \cite{DBLP:books/daglib/0097931}.

\begin{definition}[Boolean function]
\label{def:boolean_function}
A \emph{Boolean function} is a function $f: \mathbb{B}^n \rightarrow \mathbb{B}$ for some $n \in \mathbb{N}$.
\end{definition}

\begin{definition}[Family of Boolean functions]
\label{def:family_boolean_function}
A \emph{family of Boolean functions} is a sequence $f = (f_n)_{n \in \mathbb{N}}$, where $f_n$ is an $n$-ary Boolean function.
\end{definition}

\begin{definition}[Basis]
\label{def:basis}
A \emph{basis} is a finite set consisting of Boolean functions and families of Boolean functions.
\end{definition}

Informally, a \emph{Boolean circuit} is a directed acyclic graph with internal nodes marked by elements of $\{ \land , \lor, \neg \}.$
Nodes with no in-going edges are called \emph{input nodes}, and nodes with no outgoing edges are called \emph{output nodes}.
A node marked $\neg$ may have only one outgoing edge.
Computation in the circuit begins with placing input bits on the input nodes (one bit per node) and proceeds as follows.
If the outgoing edges of a node (of in-degree $d$) marked $\land$ (similarly for nodes marked $\lor$ and $\neg$) have values $v_1, v_2, ..., v_d$ then the node is assigned the value $\land^{d}_{i = 1} v_i.$
The output of the circuit is read from its output nodes.
The \emph{size} of a circuit is the number of its edges.
A \emph{polynomial-size circuit family} is an infinite sequence of Boolean circuits $\pi_1, \pi_2, ...$ such that, for every $n$, the circuit $\pi_n$ has $n$ input nodes and size $p(n)$, where $p$ is a polynomial fixed for the entire family.

\begin{definition}[Circuit]
\label{def:circuit}
Let $B$ be a basis. A \emph{Boolean circuit} over $B$ with $n$ inputs and $m$ outputs is a tuple
\[ \pi = (V, E, \alpha, \beta, \omega) ,\]
where $(V, E)$ is a finite directed acyclic graph, $\alpha : E \rightarrow \mathbb{N}$ is an injective function, $\beta : V \rightarrow B \cup \{ x_1, x_2, ..., x_n\}$, and $\omega : V \rightarrow \{ y_1, y_2, ..., y_m \} \cup \{ * \}$, such that the following conditions hold:
\begin{enumerate}
\item If $v \in V$ has in-degree 0, then $\beta(v) \in \{ x_1, x_2, ..., x_n \}$ or $\beta (v)$ is a 0-ary Boolean function (i.e.\ a Boolean constant) from $B$.
\item If $v \in V$ has in-degree $k > 0$, then $\beta (v)$ is a $k$-ary Boolean function from $B$ or a family of Boolean functions from $B$.
\item For every $i$, $1 \leq i \leq n$, there is at most one node $v \in V$ such that $\beta (v) = x_i$.
\item For every $i$, $1 \leq i \leq m$, there is at most one node $v \in V$ such that $\omega (v) = y_i$.
\end{enumerate}
\end{definition}

\begin{remark}
A Boolean circuit $\pi$ with $n$ inputs and $m$ outputs computes a Boolean function
\[ f: \mathbb{B}^n \rightarrow \mathbb{B}^m .\]
\end{remark}

\begin{definition}[Circuit family]
\label{def:circuit_family}
Let $B$ be a  basis. A \emph{circuit family} over $B$ is a sequence $\Pi = (\pi_0, \pi_1, \pi_2, ...),$ where for every $n \in \mathbb{N}$, $\pi_n$ is a circuit over $B$ with $n$ inputs.
Let $f_n$ be the function computed by $\pi_n$.
Then we say that $\Pi$ computes the function $f : \mathbb{B}^{*} \rightarrow \mathbb{B}^{*},$ defined for every $w \in \mathbb{B}^{*}$ by
\[ f(w) \overset{\text{def}}{=} f_{|w|}(w).\]
\end{definition}

\begin{remark}
For simplicity of presentation, we often abuse our notation slightly by considering circuit families $(\pi_n)_{n \in \mathbb{N}}$, where $\pi_n$ has $p(n)$ rather than $n$ input bits, for some fixed polynomial $p$.
\end{remark}

\section{Algorithms}
\label{apx:alg}

\subsection{Biometric matching}

Algorithm~\ref{alg:f} implements the function $f$ defined in \S\ref{subsec:protocol}.
\begin{algorithm}
\label{alg:f}
\Input{$\mathbf{x}, \mathbf{y} \in (\mathbb{Z}_m)^n $ and $\mathcal{B} \in \mathbb{Z}$}
\Output{$b \in \mathbb{B}$}
$ \Delta_{x,y} \leftarrow \sum_{i=1}^n (y_i - x_i)^2 $\;
\eIf{$\Delta_{x,y} \leq \mathcal{B}$}{$b=1$}{$b=0$}
\KwRet{$b$}
\caption{Pseudo-code of the biometric matching $f$}
\end{algorithm}

\subsection{Basic operations}
\label{subsec:routines}

As stated by Crawford et al.\ \cite{DBLP:conf/ccs/CrawfordGHPS18}, a key step for practical
homomorphic encryption is to implement basic routines and tools, e.g. binary arithmetic, and make
them available for use and optimisation. We implemented the following basic arithmetic functions,
needed to calculate Euclidean distance (see \S\ref{subsec:biometric_match}). In each case
pseudo-code (using mainly logic) is provided below. Apart from the specified functions, we also
used the bitwise routines implemented in the TFHE library\footnote{A list is given at:
\url{https://tfhe.github.io/tfhe/gate-bootstrapping-api.html}}. All the functions are presented as
they are executed in the plaintext domain, although the implementations of those routines are
specific to the ciphertext domain.

\emph{1-bit addition}

We denote naive binary addition by $\mathtt{1bit\_add}$\@. Two bits $a$ and $b$ are
$\mathtt{XOR}$-ed with carry; the carry is updated and returned for use in another 1-bit addition
as part of $n$-bit addition.
Algorithm~\ref{alg:1-bit-add} implements the 1-bit addition routine.

\begin{algorithm}[htb]
\label{alg:1-bit-add}
\Input{$a, b, carry_{in} \in \mathbb{B}$}
\Output{$res,  carry_{out} \in \mathbb{B}$}
$res \leftarrow a \mathtt{~XOR~} b \mathtt{~XOR~} carry_{in} $\;
$ carry_{out} \leftarrow (a \mathtt{~AND~} b) \mathtt{~XOR~} (a \mathtt{~AND~} carry_{in}) \mathtt{~XOR~} $ \\ $  (b \mathtt{~AND~} carry_{in})$\;
\KwRet{$(res, carry_{out})$}
\caption{Pseudo-code of 1-bit addition}
\end{algorithm}

\emph{\textit{n}-bit addition}

We denote naive bitwise addition by $\mathtt{nbit\_add}$\@. This routine uses $\mathtt{1bit\_add}$
and applies to all bits of two $n$-bit numbers.
Algorithm~\ref{alg:n-bit-add} implements the $n$-bit addition routine.

\begin{algorithm}[htb]
\label{alg:n-bit-add}
\Input{$a, b \in \mathbb{B}^n$}
\Output{$res \in \mathbb{B}^{n+1}$}
$carry \in \mathbb{B}$\;
$ carry \leftarrow 0$\;
\For{$i \leftarrow 1$ \KwTo $n$}{$ (res, carry) \leftarrow \mathtt{1bit\_add}(a_i, b_i, carry)$}
\KwRet{$res$}
\caption{Pseudo-code of n-bit addition}
\end{algorithm}

\emph{Two's complement}

We implemented subtraction as addition between a number and the two's complement of the other
number. Thus, we require this subroutine. We denote by $\mathring{a}$ the two's complement of $a$\
and by $\mathtt{twos}$ the two's complement function.
Algorithm~\ref{alg:twoscomp} implements the two's complement routine.

\begin{algorithm}[htb]
\label{alg:twoscomp}
\Input{$a \in \mathbb{B}^n$}
\Output{$\mathring{a} \in \mathbb{B}^{n+1}$}
\For{$ i \leftarrow 1$ \KwTo $n$}{$\mathring{a}_i \leftarrow a_i \mathtt{~XOR~} 1  $}
$\mathring{a} \leftarrow \mathtt{nbit\_add}(\mathring{a}, 1)$\;
\KwRet{$\mathring{a}$}
\caption{Pseudo-code of two's complement}
\end{algorithm}

\emph{Absolute value} 

The absolute value was required when calculating the Manhattan distance (see \S\ref{subsec:biometric_match}).
We denote this function by $\mathtt{abs}$\@.
MSB$(a)$ below outputs the Most Significant Bit of $a$.
Algorithm~\ref{alg:abs} implements the absolute value routine.

\begin{algorithm}
\label{alg:abs}
\Input{$a\in \mathbb{B}^n$}
\Output{$|a| \in \mathbb{B}^{n}$}
$mask, tmp \in \mathbb{B}^{n}$\;
\For{$ i \leftarrow 1$ \KwTo $n$}{$|mask|_i \leftarrow \textnormal{MSB}(a) $}
$tmp \leftarrow \mathtt{nbit\_add} (a, mask)$\;
\For{$ i \leftarrow 1$ \KwTo $n$}{$|a|_i \leftarrow tmp_i \mathtt{~XOR~} mask_i $}
\KwRet{$|a|$}
\caption{Pseudo-code of absolute value}
\end{algorithm}

\emph{\textit{n}-bit subtraction}

As explained above, when subtracting $b$ from $a$, the routine adds $a$ to $\mathring{b}$\@.
We denote this routine by $\mathtt{sub}$\@.
After the addition, if the final carry denoted $carry_f$ over the sum is 1, the result is positive and remains unchanged.
If not (i.e.\ $carry_f$ is 0), the result is negative thus its two's complement is returned.
The cost of a branching condition being too great, a sequence of instructions is used instead, leading to the genuine result.
Algorithm~\ref{alg:sub} implements the subtraction routine.

\begin{algorithm}
\label{alg:sub}
\Input{$ a, b \in \mathbb{B}^n$}
\Output{$res, tmp, var \in \mathbb{B}^{n+1}$}
$\mathring{b} \leftarrow \mathtt{twos}(b)$\;
$ tmp \leftarrow \mathtt{nbit\_add}(a, \mathring{b})$\;
\For{$ i \leftarrow 1$ \KwTo $n+1$}{$var_i \leftarrow carry_f $}
$res \leftarrow  \mathtt{twos}(tmp \mathtt{~AND~} \overline{var}) \mathtt{~OR~}  (tmp \mathtt{~AND~} var) $\;
\KwRet{$res$}
\caption{Pseudo-code of n-bit subtraction}
\end{algorithm}

\emph{Multiplication}

We implemented a naive multiplication algorithm; however, other algorithms have smaller complexity, e.g.\ Karatsuba multiplication \cite{Karatsuba}. Implementing this is left for future work.
Algorithm~\ref{alg:mult} implements this routine denoted by $\mathtt{mult}$\@.

\begin{algorithm}[htb]
\label{alg:mult}
\Input{$ a, b \in \mathbb{B}^n$}
\Output{$res \in \mathbb{B}^{2n}$}
$res \leftarrow 0^{2n}$\;
$tmp \leftarrow 0^{2n}$\;
\For{$ i \leftarrow 1$ \KwTo $n$}{
\For{$j \leftarrow 1$  \KwTo $n$}{ $tmp_{i+j} \leftarrow a_j \mathtt{~AND~} b_i$}
 $ res \leftarrow \mathtt{nbit\_add}(res, tmp)$ }
\KwRet{$res$}
\caption{Pseudo-code of n-bit multiplication}
\end{algorithm}

\emph{1-bit comparison}

Secure integer comparison has been studied for a long time \cite{DBLP:conf/ctrsa/BourseST20}. The
first solution was probably that of Yao \cite{DBLP:conf/focs/Yao82b} through the Millionaires'
problem. Integer comparison is very costly in terms of computation when using FHE; this is why it
is usually better to avoid computing such an operation. Moreover it can also be difficult to
articulate in ciphertext spaces. In TFHE, this operation is done using logic gates, and a proposal
for implementation is published in the tutorial section in \cite{TFHE}. The authors use a
$\mathtt{MUX}$ gate in their function, which is exhaustively explained in
\cite[Section~3.4]{DBLP:journals/joc/ChillottiGGI20}. The authors provide two functions, one to
compare bitwise and the other to compare two binary numbers, denoted by $\mathtt{1bit\_comp}$ and
$\mathtt{nbit\_comp}$, respectively. We adapted their function in our implementation.
Algorithm~\ref{alg:1bit_comp} implements the 1-bit comparison routine.

\begin{algorithm}[htb]
\label{alg:1bit_comp}
\Input{$a, b, carry \in \mathbb{B}$}
\Output{$res \in \mathbb{B}$}
$res \leftarrow  \mathtt{~MUX~}(a \mathtt{~XNOR~} b,  carry, a)$\;
\KwRet{$res$}
\caption{Pseudo-code of 1-bit comparison}
\end{algorithm}

\emph{\textit{n}-bit comparison}

This routine performs a comparison of two $n$-bit numbers using the previous routine.
Algorithm~\ref{alg:nbit-comp} implements the $n$-bit comparison routine.

\begin{algorithm}[htb]
\label{alg:nbit-comp}
\Input{$a, b \in \mathbb{B}^n$}
\Output{$res \leftarrow a?b:carry$}
$carry, tmp \in \mathbb{B}$\;
$carry \leftarrow 0$\;
\For{$i \leftarrow 1$  \KwTo $n$}{$tmp \leftarrow\mathtt{1bit\_comp}(a_i, b_i, carry)$}
\For{$i \leftarrow 1$  \KwTo $n$}{$res \leftarrow \mathtt{MUX}(carry, b_i, a_i)$}
\KwRet{$res$}
\caption{Pseudo-code of $n$-bit comparison}
\end{algorithm}

%
%
%
%
%

\end{document}